# Electric field induced magnetic domain wall tilting

A.P. Pyatakov[1] *, A.S. Sergeev[1], D.A. Sechin[1], A.V. Nikolaev[1], E.P. Nikolaeva[1], L.E. Calvet[2]

1) Physics Department M.V. Lomonosov MSU, Moscow
2) Institut d'Electronique Fondamentale, CNRS, France

*) pyatakov@physics.msu.ru

The inclination of the magnetic domain wall plane in electric field is observed. The simple theoretical model that takes into account the spin flexoelectricity is proposed. The value of electric polarization of the magnetic domain wall was estimated as $0.3\mu C/m^2$ that agrees well with the results of electric field driven magnetic domain wall motion measurements.

In the series of theoretical works [1-5] it was predicted that magnetic inhomogeneites locally reducing the symmetry of the crystal give rise to the electrically polarized areas even in centrosymmetrical media (spin flexoelectric effect). Later this hypothesis was experimentally proved in series magnetooptical studies of micromagnetic structure in iron garnet films [6-10]. The domain walls either attract or repel from the tip electrode depending on the electric polarity of the probe and the chirality of the domain wall (fig.1) in accordance with the theory.

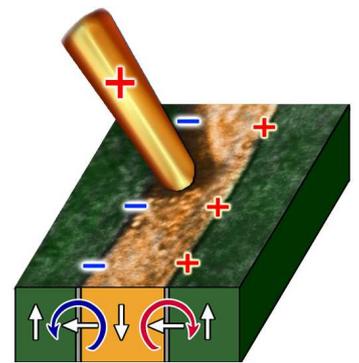

Fig. 1 Electric polarization of the domain walls: the chirality of the domain wall determines the result of interaction with the charged probe.

Meanwhile the point charge geometry is not the only electric field configuration that makes it possible to observe electric field induced changes of the micromagnetic structure. In this short paper we report on the observation of the influence of an electric field from a stripe electrode on a magnetic domain wall.

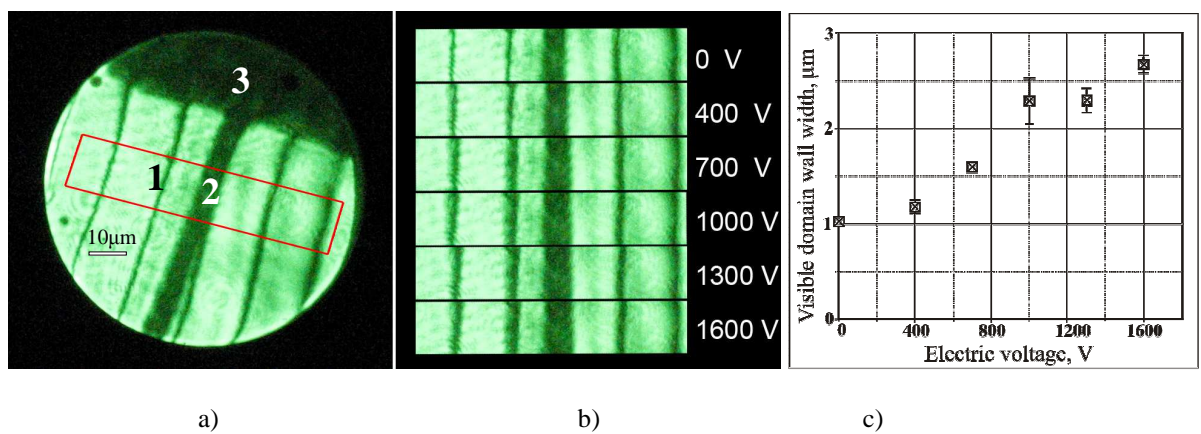

Fig.2. Domain wall image broadening under the influence of electric field a) magneto-optical images of the magnetic domain structure under the influence of electric field from a stripe electrode: 1 is the domain wall, 2 is a stripe electrode, 3 is a contact pad. b) the selected area of figure (a) at various stripe electrode voltages   c) the dependence of the apparent width on the voltage.

An iron garnet film $(BiLu)_3(FeGa)_5O_{12}$ grown epitaxially on a (210) substrate [11] was used (the film thickness was 18.7 μm, the stripe domain period was 26 μm, and the magnetization $M_s$ =5 G). A voltage was applied between the contact pad and the substrate, the micromagnetic structure transformation was observed magneto-optically in the Faraday mode (fig.2).

A broadening of the images of the domain walls oriented parallel to the stripe electrodes is observed. It increases monotonically up to the three times as the voltage grows from 0 to 1600 V (fig.2). This change of the apparent width of the domain wall can be attributed to the tilt of the domain wall plane by an angle ~10° with respect to the normal to the film (fig. 3).

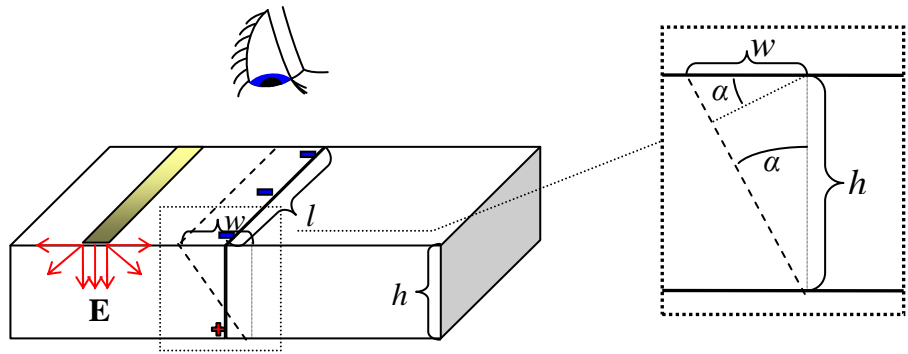

Fig.3 Schematic representation of the problem under consideration: electric field E of the positively charged electrode acts on the electric dipole of the domain wall (the corresponding bound surface charges are shown with «+» and «-»), *h* is the film thickness, *l* is the fragment of the domain wall, *w* is apparent width of the domain wall, *α* is the tilting angle of the domain wall, the dashed line is designates the inclined domain wall. The bending is exaggerated for illustrative purposes.

In accordance with the theory of spin flexoelectricity the domain wall is an electric dipole that tends to orient along the electric field (fig. 3). The tilt angle is the result of the competition between the electrostatic energy and the surface energy of the domain wall. As it is shown in the figure 3, the increment of the domain wall surface area is $\Delta S \approx l\alpha w$, where *l* is the length of the domain wall fragment, *w* is apparent width of the domain wall, and the tilt angle $\alpha \approx w/h$, where *h* is the magnetic film thickness. The energy growth of the energy of the wall with surface tension σ is paid by the exceeding drop of the electrostatic energy $\alpha P_S E l h \Delta$, where $P_S$ is the spontaneous polarization of the domain wall, *E* is the electric field, Δ is the real domain wall width resulting from the competition between the magnetic exchange and magnetic anisotropy. Thus minimization of the energy leads to the equation for domain wall tilt:

$$\alpha \sigma = P_S E \Delta$$

Using the experimental values $\alpha=10°$, $\sigma=0.01$ erg/cm$^2$, E~600kV/cm $\approx 2\cdot 10^3$ CGS, $\Delta=10^{-5}$ cm, we obtain an estimate for electric polarization $P_S$~ 0.1 CGS = 0.3 μC/m$^2$ that agrees well with the estimate $P_S=M_S H/E$, obtained from a domain wall dynamics measurements [7] in magnetic field H=50 Oe that accelerate the domain wall to the same velocity as an electric field E ~ 1MV/cm $\approx 3\cdot 10^3$ CGS (the magnetization of the film is $M_S\approx$5Gs).